# Discovery of a hybridization-wave electronic order in a van der Waals Kondo lattice

Lu Cao[1,2], Jiefei Shi[1], Lanxin Liu[4,5], Xuan Luo[4], Yu-Ping Sun[6,4,7], Yi-feng Yang[3,2], Yugui Yao[8*], Jinhai Mao[2*], Yuhang Jiang[1*]

[1] College of Materials Science and Optoelectronic Technology, Center of Materials Science and Optoelectronic Engineering, University of Chinese Academy of Sciences, Beijing 100049, China.
[2] School of Physical Sciences, University of Chinese Academy of Sciences, Beijing 100049, China.
[3] Beijing National Laboratory for Condensed Matter Physics and Institute of Physics, Chinese Academy of Sciences, Beijing, 100190, China.
[4] Key Laboratory of Materials Physics, Institute of Solid State Physics, HFIPS, Chinese Academy of Sciences, Hefei 230031, China
[5] University of Science and Technology of China, Hefei 230026, China
[6] Anhui Key Laboratory of Low-Energy Quantum Materials and Devices, High Magnetic Field Laboratory, HFIPS, Chinese Academy of Sciences, Hefei 230031, China
[7] Collaborative Innovation Center of Advanced Microstructures, Nanjing University, Nanjing 210093, China
[8] Centre for Quantum Physics, Key Laboratory of Advanced Optoelectronic Quantum Architecture and Measurement (MOE), School of Physics, Beijing Institute of Technology, Beijing, 100081, China.

*Correspondence to: yuhangjiang@ucas.ac.cn, jhmao@ucas.ac.cn, ygyao@bit.edu.cn

**Abstract**

Kondo lattice systems, in which localized magnetic moments coherently hybridize with itinerant electrons, exhibit a rich landscape of emergent quantum phenomena. Within this framework, the hybridization strength itself has been theoretically proposed as a spatially modulated order parameter, giving rise to a so-called '*hybridization wave*'. However, direct experimental evidence of this quantum state has remained an outstanding challenge. Here, we report the direct observation of a hybridization wave in the layered transition metal dichalcogenide *6R*-TaS$_2$, a naturally occurring heterostructure composed of alternating 1*T*- and 1*H*-TaS$_2$ layers. Using scanning tunneling microscopy and spectroscopy (STM/STS), we identify the hybridization gap $\Delta_{\text{hyb}}$ in 1*T* layer, demonstrating the establishment of a coherent Kondo lattice. Notably, we discover that the hybridization gap present a uniaxial unit-cell doubling modulation, which breaks the both translational and rotational symmetries of the underlying Star-of-David superlattice. Such unit-cell doubling is not caused by structural topography, and therefore, constitutes the real-space visualization of the hybridization-wave order. Furthermore, the hybridization wave correlates with an energy-dependent nematic order that shares the same periodicity and orientation, revealing intertwined electronic instabilities. Our findings not only validate a long-standing prediction but also establish layer-engineered van der Waals materials as a versatile platform for exploring and controlling hybridization-driven quantum phases.

Strongly correlated electron systems often defy description within conventional paradigms of solid-state physics. This gives rise to emergent states such as heavy fermion behavior[1], unconventional superconductivity[2], non-Fermi liquid states[3] and quantum criticality[4]. A key element in many of these phenomena is Kondo lattice[5], where a periodic array of localized magnetic moments hybridizes coherently with itinerant electrons. This hybridization renormalizes the electronic structures, creating heavy fermion bands near the Fermi level ($E_F$) and characteristic hybridization gaps[6]. While the single-impurity Kondo effect has been well established[7-9], its lattice counterpart is still less understood. The coherent hybridization enhances the effective quasiparticle mass and allows for the emergence of novel ordered states that are not accessible in the impurity limit. Among the most difficult to grasp is the *hybridization wave*[10-12], a theoretically predicted state in which the hybridization strength itself varies spatially. This state has been proposed as a potential explanation for the hidden order in heavy fermion compounds[10,13-15], yet direct spectroscopic evidence has been lacking. Proving the existence of a hybridization wave not only confirms that hybridization strength can act as an order parameter capable of breaking translational or rotational symmetry, but also offers a new perspective for understanding instabilities in Kondo lattice systems. The lack of suitable platforms, however, has impeded experimental progress.

Layered transition metal dichalcogenides (TMDs) accommodate a diverse range of quantum states, including charge density wave[16-19] (CDW), superconductivity[20-24] and strongly correlated electronic states[25,26]. The van der Waals (vdW) stacking of TMDs allows for the formation of heterostructures with distinct electronic characters in different layers[27-31]. These configurations offer the promising platforms for engineering exotic phenomena resulting from interlayer coupling, such as topological superconductivity[27] and artificial heavy fermion[28]. Among TMDs, $6R$-TaS$_2$ features a rhombohedral stacking of alternating $1T$- and $1H$-TaS$_2$ layers[32,33]. The $1T$ layers possess localized magnetic moments associated with Star-of-David (SD) superlattice and are regarded as a candidate for quantum spin liquid[34,35]. In contrast, the $1H$ layers are metallic and supply itinerant electrons[28]. This architecture inherently realizes a two-dimensional Kondo lattice at the van der Waals scale[28,30,36], presenting unprecedented opportunities to probe hybridization physics with atomic resolution.

In this work, we report the discovery of the long-sought hybridization wave in $6R$-TaS$_2$. Scanning tunneling microscopy/spectroscopy (STM/STS) shows the hybridization gap $\Delta_{hyb}$ associated with the $1T$ layers, thus confirming the formation of a Kondo lattice and heavy fermion behavior[6,37-40]. Subsequently, we disclose the nanoscale oscillations of $\Delta_{hyb}$ that are related to the SD superlattice period $a_{SD}$. Most importantly, we further discover a uniaxial unit-cell doubling of $\Delta_{hyb}$, breaking the translational and rotational symmetry of the SD superlattice. This doubling, which is not present in the topography, serves as spectroscopic evidence of a *hybridization wave*[10]. Moreover, we observe an energy-dependent nematic order that aligns with the hybridization wave, highlighting the close coupling between these two phenomena. Our findings not only provide a direct visualization of a hybridization wave but also put forward $6R$-TaS$_2$ as a prototypical platform for engineering and probing novel hybridization-driven orders in two-dimensional vdW systems.

## Results

### Kondo lattice and hybridization gap in $6R$-TaS$_2$

We begin by characterizing the electronic structure of $6R$-TaS$_2$. The lattice structure of

$6R$-TaS$_2$ is displayed in Fig. 1a. Along the crystallographic $c$-axis, the $1T$-TaS$_2$ and $1H$-TaS$_2$ layers stack alternately. Every two $1T/1H$ bilayers (marked by red rectangles in Fig. 1a) is related by an in-plane glide symmetry, as indicated by the red dashed lines in Fig. 1a. A full stacking cycle consists of six layers (black dashed rectangle in Fig. 1a), forming the rhombohedral structure ($6R$). This alternate stacking gives rise to two distinct surface terminations after cleavage, corresponding to the $1T$ and $1H$ surfaces, respectively. On the $1T$ terminal surface, the typical STM topographic image exhibits a $\sqrt{13} \times \sqrt{13}$ SD superlattice (Fig. 1b) with a period $a_{SD} = 1.18$ nm (Supplementary Fig. 1a-c), consistent with previous results on monolayer and bulk $1T$-TaS$_2$ (ref. [28,41]). On the $1H$ terminal surface, the representative STM topography shows clear atomic resolution with a period $a_0 = 0.33$ nm (Supplementary Fig. 1d-f). Additionally, a $3 \times 3$ CDW pattern is observed[42], confirmed by its fast Fourier transform (FFT) (Supplementary Fig. 1f). These structural disparities enable us to easily identify them in the subsequent studies.

The mechanism for the formation of the Kondo lattice in $6R$-TaS$_2$ is as follows. In the $1T$ layers, each SD unit-cell contributes an unpaired $d$ electron, which carries a local magnetic moment[34]. Due to the characteristic of the $d$-orbital, these local moments are not confined at single sites but are extended within the entire SD unit-cells[43,44], forming an array defined by the SD superlattice. On the other hand, the $1H$ layers exhibit the metallic behavior and provide itinerant electrons. When the $1T$ and $1H$ layers stack together, the Kondo coupling between the local magnetic moment array and itinerant electrons (Fig. 1c) leads to the formation of a Kondo lattice and the heavy fermion state, with hybridization bands emerging near $E_F$ (solid line in Fig. 1d). This scenario is demonstrated by recent experimental observations on molecular-beam-epitaxy grown $1T/1H$ TaS$_2$ heterostructure[28]. Consequently, there should be the hybridization gap $\Delta_{hyb}$ in reciprocal-space[15], as sketched in Fig. 1d.

The hybridization gap $\Delta_{hyb}$ can be probed through the tunneling spectrum (d$I$/d$V$). When the tunneling into the local orbital is dominant, the spectrum presents two distinct hybridization peaks that reflect the hybridization gap $\Delta_{hyb}$ (Fig. 1e), as previously demonstrated in ref [15]. This is exactly the spectral feature observed on the $1T$ surface (Fig. 1f), where a significant tunneling amplitude into the local orbital occurs[31]. Henceforth, the two hybridization peaks flanking the $E_F$ are denoted as P$_1$ and P$_2$ (Fig. 1f). To affirm the Kondo coupling nature of the hybridization peaks P$_1$ and P$_2$, we conduct additional magnetic field dependent measurements. We observe a clear Zeeman splitting behavior of the hybridization peak under high magnetic fields (Supplementary Fig. 2), demonstrating the Kondo coupling scenario[30]. These hybridization peaks are consistent with previous theoretical findings[6,38] and STS measurements[15,31,36] and the observation of the hybridization gap provides evidence of a coherent Kondo lattice in the $6R$-TaS$_2$ system.

**Temperature and Spatial evolution of hybridization peaks**

The temperature dependence of the d$I$/d$V$ spectra collected from the $1T$-TaS$_2$ surface are plotted in Fig. 2a, acquired from 0.3 K to 20 K. As temperature rises, both peaks P$_1$ and P$_2$ weaken and broaden. To quantitatively characterize this thermal effect, we extract the full width at half maximum (FWHM) for P$_1$ and P$_2$ through multiple-peak fitting as reported in other Kondo lattice system[15] (see Methods), and display the results in Fig. 2b. The FWHM for P$_2$ is always larger than that of P$_1$, implying a particle-hole asymmetry of the hybridization band. In addition, the spectrum shows pronounced spatial dependence. For instance, the d$I$/d$V$ spectra measured at SD center (Fig. 2c) and edge (Fig. 2d) reveal distinct peak positions. Specifically, $P_1$ appears at about –1.3 mV at SD center (Fig. 2c) and –0.7 mV at edge (Fig. 2d), while $P_2$ locates at approximately +3.5 mV at SD center (Fig. 2c) and +4.7 mV at the edge (Fig. 2d). To be more accurately, we utilize multiple-peak fitting[15] to extract the energy locations ($E_{P1}$ and $E_{P2}$) for these two peaks (see Methods) and then calculate the hybridization gap size by $\Delta_{hyb} = E_{P2} - E_{P1}$. The results yield $\Delta_{hyb}$ = 4.81 meV at the SD center and $\Delta_{hyb}$ = 5.41 meV at the edge. This spatial variation already suggests a nontrivial electronic modulation associated with the SD superlattice. More examples of the fitting results at different spatial locations are shown in Supplementary Fig. 3a-c.

To obtain the detailed spatial evolution of $\Delta_{hyb}$, we acquired one-dimensional spectra map d$I$/d$V$(**r**, $V$) at region #1 (Fig. 2e) following a high symmetric direction, **d**$_1$, of the SD superlattice (orange arrow in Fig. 2e). The map starts from an SD center and traverses five SD periods, with the spectra displayed as a water-fall plot in Fig. 2f. The energy locations of $P_1$ and $P_2$ exhibit clear spatial oscillations, as indicated by the dashed lines. To quantify this modulation, we extracted $E_{P1}$(**r**) and $E_{P2}$(**r**) (Supplementary Fig. 3d-e) and calculated the spatial dependence of the hybridization gap, $\Delta_{hyb}$(**r**). In Fig. 2g, we plot the topographic height $T$(**r**) (upper panel) and $\Delta_{hyb}$(**r**) (lower panel) for direct visual comparison. We find that $\Delta_{hyb}$ reaches a local minimum at the SD center and increases at the SD edge. Overall, $\Delta_{hyb}$(**r**) oscillates with the same period as the SD superlattice ($a_{SD}$). FFT analyses reveal dominant wave vectors at $2\pi/a_{SD}$ in both $T$(**q**) and $\Delta_{hyb}$(**q**) (Fig. 2h), further confirming that the hybridization oscillations are phase-locked to the underlying SD superlattice.

## Evidence for hybridization wave

Beyond the SD-periodic modulation, a more distinctive behavior of $\Delta_{hyb}$ emerges along the high-symmetry direction **d**$_2$ (red arrow in Fig. 2e). The d$I$/d$V$(**r**, $V$) spectra measured along **d**$_2$ reveal an alternating pattern: adjacent SD centers exhibit $P_2$ peaks at different energies (black arrows in Fig. 3a), effectively doubling the periodicity. This feature is absent in the d$I$/d$V$(**r**, $V$) spectra taken along **d**$_1$ (Fig. 2f). Using the same fitting method described above (more fitting results can be found in Supplementary Fig. 4), the extracted $T$(**r**) and $\Delta_{hyb}$(**r**) are depicted in Fig. 3b. Significant new minima of $\Delta_{hyb}$ appear at every two SD periods (marked by vertical arrows in Fig. 3b), indicating a unit-cell doubling behavior with period $2a_{SD}$. This unit-cell doubling is clearly visible in the FFT spectrum $\Delta_{hyb}$(**q**) (Fig. 3c lower panel), where a new wave vector appears at $2\pi/2a_{SD}$. In contrast, neither $T$(**r**) nor $T$(**q**) exhibits such doubling behavior (upper panels in Figs. 3b-c). The unit-cell doubling of $\Delta_{hyb}$(**r**) is also clearly observed along the third high-symmetry direction, **d**$_3$ (Supplementary Fig. 5).

To further visualize this unit-cell doubling, we conduct the two-dimensional tunneling conductance maps, d$I$/d$V$(***r***, $E$ = e$V$) ≡ $g$(***r***, $E$ = e$V$), over a dense spatial grid in region #1 and extract the spatial distribution $\Delta_{hyb}$(**r**), as shown in Fig. 3d. A uniaxial modulation, which induces the unit-cell doubling of $\Delta_{hyb}$, is clear resolved. We highlight this unit-cell doubling by solid parallelogram compared to the original SD unit-cell indicated by dashed diamond in Fig. 3d. The FFT image $\Delta_{hyb}$(**q**) (Fig.3e) reveals wave vectors at **q**$_1$, **q**$_2$ and **q**$_3$ corresponding to SD oscillations along the three high-symmetry directions, and wave vector at **q**$_4$ signifying the uniaxial unit-cell doubling modulation. This uniaxial unit-cell doubling, absent in the simultaneously acquired topography $T$(**r**) and its FFT $T$(**q**) (Fig. 3f-g), rules out a trivial topographic origin and suggests a purely electronic order that possess spontaneous spatial modulation of the hybridization strength itself.

Note that the hybridization strength is related to the FWHM of hybridization peak [45]. Accordingly, we extract the FWHM maps for peaks $P_1$ and $P_2$ (denoted as $FWHM_{P1}(\mathbf{r})$ and $FWHM_{P2}(\mathbf{r})$), as shown in Figs. 3h-i. Both $FWHM_{P1}(\mathbf{r})$ and $FWHM_{P2}(\mathbf{r})$ exhibit the same uniaxial unit-cell doubling features as $\Delta_{hyb}(\mathbf{r})$ (highlighted by solid parallelogram in Figs. 3h-i), which are further confirmed by the wave vectors in the corresponding FFT images (red dashed circles in the insets of Figs. 3h-i). The one-dimensional character of these FWHM maps demonstrates that the hybridization strength in the Kondo lattice is indeed spatially modulated in a unidirectional manner.

In conventional Kondo lattice models, the hybridization is considered to be onsite. Therefore, the oscillation of $\Delta_{hyb}$ observed here, involving an extended local-moment orbital, is unexpected and has not been reported in other Kondo lattice systems. Most importantly, the additional unit-cell doubling of $\Delta_{hyb}$ breaks the translational and rotational symmetry and indicates the emergence of a spontaneously modulated hybridization order, i.e., a *hybridization wave*[10,12]. This constitutes the direct spectroscopic evidence for a hybridization wave in a Kondo lattice system. Notably, the hybridization wave is well reproduced at other regions. In Supplementary Figs. 6-7, we present an additional reproducible dataset in region #2, in which the uniaxial hybridization wave is along a different direction.

**Energy-dependent nematic order**

Interestingly, tunneling conductance maps at fixed energies reveal an additional nematic order. For example, in map $g(\mathbf{r}, -1 \text{ meV})$ (Fig. 4a), the one-dimensional stripe features appear along each row of SD superlattice, with alternating LDOS intensity between adjacent rows. Figure 4b is the line-profile extracted from Fig. 4a, with the spatial locations for relatively high and low LDOS intensity marked by the red and green arrows, respectively. The results in Fig. 4a-b demonstrate the electronic anisotropy caused by the nematic order, which breaks the translational and rotational symmetry of the SD superlattice. The FFT image $g(\mathbf{q}, -1 \text{ meV})$ confirms nematic order along a preferred axis (red dashed circles in Fig. 4c). We note the nematic order shares the same orientation and periodicity as the hybridization wave, suggesting an intertwined electronic instability.

The nematic order also exhibits a striking energy dependence. We compare the LDOS distributions measured at $E = -1$ meV (Fig. 4a-c), 0 meV (Fig. 4d-f) and +2 meV (Fig. 4g-i). At $E = -1$ meV, $g(\mathbf{r}, -1 \text{ meV})$ shows pronounced uniaxial nematicity. At 0 meV, the nematic signal vanishes in both real-space (Fig. 4d-e) and reciprocal-space (Fig. 4f). At $E = +2$ meV, the nematic order reappears (Fig. 4g-i). However, the spatial locations for relatively high and low LDOS intensity are interchanged compared with that at $E = -1$ meV (arrows in Fig. 4b,h). Therefore, the nematic order is energy-dependent. Furthermore, this nematic order remains unchanged under in-plane ($B_x = 1$ T) and out-of-plane ($B_z = 3$ T and 6 T) magnetic fields (Supplementary Fig. 8). The lack of response to the magnetic fields excludes the possibility of a magnetic ground state for this nematic order.

An important question is whether this nematic order originates from a trivial CDW. First, a CDW phase typically manifests as a periodic modulation in the surface topography[46,47]. However, only the SD superlattice is observed in the topography of the hybridization regions. Second, a CDW phase is usually accompanied by a gap feature near $E_F$ that is readily detectable by tunneling spectroscopy[47-49]. No such gap is observed in our measurements. Third, a CDW phase typically induces LDOS modulations that persist across all energy levels, including at $E_F$ (ref. [50]). This behavior contrasts with the energy-dependent nematic order, which vanishes at $E_F$ (Fig. 4d-f). Therefore, we conclude the nematic order is unrelated to a CDW phase. The shared orientation and periodicity between the hybridization wave and the nematic order indicate an intrinsic nematic instability coupled to the hybridization wave. However, determining which is the precursor remains an open question.

**Discussion**

Our observations provide direct evidence for the existence of hybridization waves in a Kondo lattice. The observed unit-cell doubling in $\Delta_{hyb}$ is distinct from the single Kondo holes phenomenon[51-54], where the hybridization strength varies only at specific non-magnetic sites. Instead, it signifies a new form of electronic order, a spontaneously spatial-modulated hybridization strength. Moreover, the extended nature of the localized moment, which spans the entire SD unit-cell, goes beyond the conventional Kondo lattice model of atomically localized magnetic moment. Consequently, new physics may emerge, manifesting as unexpected hybridization patterns on the nanometric scales. Our findings call for further theoretical and experimental investigations into systems with extended magnetic units, such as the moiré superlattices[55-57] or other cluster-based magnets[25,30].

Equally remarkable is the coexistence of the hybridization wave with an energy-dependent nematic order, suggesting a potential coupling between these two orders. Their coexistence also emphasizes the importance of introducing more complex hybridization structures in modeling. One example is the theoretical proposed hastatic order[58-60] for heavy fermion systems. In this order, the hybridization involves integer-spin local moments and half-integer-spin itinerant electrons, forming a spinor order parameter that leads to energy-dependent nematicity[58]. While such hastatic order has not been definitively observed in three-dimensional heavy fermion compounds, the Kondo lattice in $6R$-TsS$_2$ offers a tangible and tunable van der Waals platform for exploring such ideas.

Together, our observations establish a comprehensive picture of the Kondo lattice in $6R$-TaS$_2$. The interlayer coupling forms a coherent Kondo lattice, within which the hybridization gap $\Delta_{hyb}$ exhibits spatial oscillations that are locked to the SD superlattice. Superimposed on this background is a uniaxial unit-cell doubling of $\Delta_{hyb}$, providing direct evidence for a hybridization wave. The intimate correlation between this hybridization wave and the energy-dependent nematic order reveals a rich landscape of intertwined electronic orders. Our work positions the $1T/1H$ TaS$_2$ heterostructures as a model system for exploring quantum coherence and emergent phenomena in two dimensions, with broad implications for heavy fermion physics, nematicity, and the design of van der Waals heterostructures.

## Methods

### Sample synthesis

$6R$-$TaS_2$ single crystals are synthesized by the chemical vapor transport (CVT) method[61]. Stoichiometric amounts of tantalum (99.9% purity) and sulfur (99.9%) powders are sealed in evacuated quartz tubes together with iodine as the transport agent. The tubes are placed in a two-zone furnace with a temperature gradient of 850 °C (source) to 750 °C (growth). Plate-like single crystals with typical dimensions of 2 × 2 mm² are obtained. The rhombohedral $6R$ polytype is confirmed by single-crystal X-ray diffraction[61].

### STM/S measurements

The samples which have a size about 2 × 2 mm² are mounted on sample holders by epoxy. The epoxy is cured below 80 °C to prevent any possible structure change of $6R$-$TaS_2$ lattice. The samples are cleaved at room temperature in ultra-high vacuum chamber ($< 1 \times 10^{-10}$ mbar), yielding fresh surface for measurement. We note the as-cleaved surface may be shiny or not, depending on different pieces of sample. We select samples with shiny cleaved surfaces for further measurements. After cleavage, the sample is transferred in suit into the scanning head within 2 minutes. All measurements are performed at 4.2 K except special annotation. Pt-Ir tips are used in STM/STS measurements. The topography images are obtained under constant-current mode. The $dI/dV$ spectra or maps are acquired by a standard lock-in amplifier with a modulation voltage $V_{mod}$ = 1 mV added to the sample bias. Magnetic fields are applied perpendicular or parallel to the sample surface.

### Kondo coupling nature of the hybridization peak

To identify the Kondo coupling nature of hybridization peak, we perform the magnetic field ($B$) dependence measurements under 0.3 K where a high energy-resolution is accessible. The results are displayed in Supplementary Fig. 2a. As the magnetic field increases, hybridization peak $P_1$ shows a clear splitting behavior. We use the multiple-peak with Gaussian peaks to extract the energy splitting $\Delta E$ (Supplementary Fig. 2b) and plot $\Delta E$ as a function of $B$ (Supplementary Fig. 2c). Due to the linear tendency between $\Delta E$ and $B$, we fit $\Delta E(B)$ by Zeeman splitting $\Delta E = g_{eff}\mu_B B$, where $g_{eff}$ is the effective Landé $g$-factor, $\mu_B$ is the Bohr magneton. The fitting yields $g_{eff}$ = 1.21±0.01. We note $P_2$ also gradually splits because of its broadening under B = 6~8 T (Supplementary Fig. 2a), suggesting a small $g_{eff}$. The Zeeman splitting behavior under magnetic fields supports the Kondo coupling nature of the hybridization peak.

### Multi-peak fitting for FWHM and peak energy location extraction

Due to the obvious two-peak spectral feature, we use the multi-peak fitting method to extract the energy position, $E_{P1}$ and $E_{P2}$, for the two hybridization peaks as reported in previous Kondo lattice study[15]. The $dI/dV(V)$ spectra are fit by two Gaussians peaks plus a constant background, as described by the following formula:

$$\frac{dI}{dV}(V) = \left[ a_1 \exp\left(-\frac{(eV - E_{P1})^2}{2\sigma_1^2}\right) + a_2 \exp\left(-\frac{(eV - E_{P2})^2}{2\sigma_2^2}\right) + d \right]$$

Where $V$ is the sample bias, $e$ is the elementary charge, $a_1$ and $a_2$ are the amplitude for Gaussian peaks, $E_{P1}$ and $E_{P2}$ are the energy positions for peaks $P_1$ and $P_2$, $\sigma_1$ and $\sigma_2$ are the standard deviation, $d$ is a constant background. The full width at half maximum

(FWHM) for $P_1$ ($P_2$) is $2\sqrt{2ln2}\sigma_1$ ($2\sqrt{2ln2}\sigma_2$). In Supplementary Fig. 3a-c and Supplementary Fig. 4a-c, colored dashed curves are individual Gaussian peaks and black solid curves are the fitting results, which match well with the experimental data (colored circles). The individual Gaussian peaks yield the accurate peak energy positions for $P_1$ and $P_2$. This method can also be applied to the one-dimensional or two-dimensional d$I$/d$V$ spectra maps to obtain the spatial evolutions of $E_{P1}(\mathbf{r})$ and $E_{P2}(\mathbf{r})$. Finally, the spatial evolution of hybridization wave is calculated by $\Delta_{hyb}(\mathbf{r}) = E_{P2}(\mathbf{r}) - E_{P1}(\mathbf{r})$ (Supplementary Fig. 3d-f and Supplementary Fig. 4d-f).

**Data availability**
The data that support the findings of this work are available within the main text and Supplementary Information. The data are available from the corresponding authors upon request.

**Code availability**
The code used to analyze the data in this work is available from the corresponding authors upon request.

**References**

1. Stewart, G. R. Heavy-fermion systems. *Rev. Mod. Phys.* **56**, 755-787 (1984).
2. Smidman, M. *et al.* Colloquium: Unconventional fully gapped superconductivity in the heavy-fermion metal $CeCu_2Si_2$. *Rev. Mod. Phys.* **95**, 031002 (2023).
3. Löhneysen, H. v., Rosch, A., Vojta, M. & Wölfle, P. Fermi-liquid instabilities at magnetic quantum phase transitions. *Rev. Mod. Phys.* **79**, 1015-1075 (2007).
4. Si, Q. M. & Steglich, F. Heavy Fermions and Quantum Phase Transitions. *Science* **329**, 1161-1166 (2010).
5. Doniach, S. The Kondo lattice and weak antiferromagnetism. *Physica* **91B**, 231-234 (1977).
6. Martin, R. M. Fermi-Surfae Sum Rule and its Consequences for Periodic Kondo and Mixed-Valence Systems. *Phys. Rev. Lett.* **48**, 362-365 (1982).
7. Nagaoka, K., Jamneala, T., Grobis, M. & Crommie, M. F. Temperature dependence of a single Kondo impurity. *Phys. Rev. Lett.* **88**, 077205 (2002).
8. Ternes, M., Heinrich, A. J. & Schneider, W. D. Spectroscopic manifestations of the Kondo effect on single adatoms. *J. Phys.: Condens. Matter* **21**, 053001 (2009).
9. Morr, D. K. Theory of scanning tunneling spectroscopy: from Kondo impurities to heavy fermion materials. *Rep. Prog. Phys.* **80**, 014502 (2017).
10. Dubi, Y. & Balatsky, A. V. Hybridization wave as the "hidden order" in $URu_2Si_2$. *Phys. Rev. Lett.* **106**, 086401 (2011).
11. Su, J. J., Dubi, Y., Wolfle, P. & Balatsky, A. V. A charge density wave in the hidden order state of $URu_2Si_2$. *J. Phys.: Condens. Matter* **23**, 094214 (2011).
12. Xie, N., Hu, D. & Yang, Y.-F. Hybridization oscillation in the one-dimensional Kondo-Heisenberg model with Kondo holes. *Sci. Rep.* **7**, 11924 (2017).
13. Schmidt, A. R. *et al.* Imaging the Fano lattice to 'hidden order' transition in $URu_2Si_2$. *Nature* **465**, 570-576 (2010).
14. Aynajian, P. *et al.* Visualizing the formation of the Kondo lattice and the hidden order in $URu_2Si_2$. *Proc. Natl. Acad. Sci. USA* **107**, 10383-10388 (2010).
15. Aynajian, P. *et al.* Visualizing heavy fermions emerging in a quantum critical Kondo lattice. *Nature* **486**, 201-206 (2012).
16. Cho, D. *et al.* Nanoscale manipulation of the Mott insulating state coupled to charge order in $1T\text{-}TaS_2$. *Nat. Commun.* **7**, 10453 (2016).
17. Ma, L. *et al.* A metallic mosaic phase and the origin of Mott-insulating state in $1T\text{-}TaS_2$. *Nat. Commun.* **7**, 10956 (2016).
18. Mraz, A. *et al.* Manipulation of fractionalized charge in the metastable topologically entangled state of a doped Wigner crystal. *Nat. Commun.* **14**, 8214 (2023).
19. Tilak, N. *et al.* Proximity induced charge density wave in a graphene/$1T\text{-}TaS_2$ heterostructure. *Nat. Commun.* **15**, 8056 (2024).
20. Sipos, B. *et al.* From Mott state to superconductivity in $1T\text{-}TaS_2$. *Nat. Mater.* **7**, 960-965 (2008).
21. Xi, X. *et al.* Ising pairing in superconducting $NbSe_2$ atomic layers. *Nat. Phys.* **12**, 139-143 (2015).
22. Ribak, A. *et al.* Chiral superconductivity in the alternate stacking compound $4Hb\text{-}TaS_2$. *Sci. Adv.* **6**, eaax9480 (2020).
23. Nayak, A. K. *et al.* Evidence of topological boundary modes with topological nodal-point superconductivity. *Nat. Phys.* **17**, 1413-1419 (2021).

24. Wan, Z. *et al.* Unconventional superconductivity in chiral molecule-TaS$_2$ hybrid superlattices. *Nature* **632**, 69-74 (2024).
25. Liu, M. K. *et al.* Monolayer 1T-NbSe$_2$ as a 2D-correlated magnetic insulator. *Sci. Adv.* **7**, eabi6339 (2021).
26. Fei, Y., Wu, Z., Zhang, W. & Yin, Y. Understanding the Mott insulating state in 1*T*-TaS$_2$ and 1*T*-TaSe$_2$. *AAPPS Bull.* **32** (2022).
27. Kezilebieke, S. *et al.* Topological superconductivity in a van der Waals heterostructure. *Nature* **588**, 424-428 (2020).
28. Vano, V. *et al.* Artificial heavy fermions in a van der Waals heterostructure. *Nature* **599**, 582-586 (2021).
29. Shen, S. *et al.* Coexistence of Quasi-two-dimensional Superconductivity and Tunable Kondo Lattice in a van der Waals Superconductor. *Chin. Phys. Lett.* **39**, 077401 (2022).
30. Wan, W. *et al.* Evidence for ground state coherence in a two-dimensional Kondo lattice. *Nat. Commun.* **14**, 7005 (2023).
31. Ayani, C. G. *et al.* Probing the Phase Transition to a Coherent 2D Kondo Lattice. *Small* **20**, e2303275 (2024).
32. Achari, A. *et al.* Alternating Superconducting and Charge Density Wave Monolayers within Bulk 6R-TaS$_2$. *Nano Lett.* **22**, 6268-6275 (2022).
33. Liu, S. B. *et al.* Nematic Ising superconductivity with hidden magnetism in few-layer 6*R*-TaS$_2$. *Nat. Commun.* **15**, 7569 (2024).
34. Law, K. T. & Lee, P. A. 1T-TaS$_2$ as a quantum spin liquid. *Proc. Natl. Acad. Sci. USA* **114**, 6996-7000 (2017).
35. Chen, H. *et al.* Spectroscopic Evidence for Possible Quantum Spin Liquid Behavior in a Two-Dimensional Mott Insulator. *Phys. Rev. Lett.* **134**, 066402 (2025).
36. Ayani, C. G. *et al.* Electron delocalization in a 2D Mott insulator. *Nat. Commun.* **15**, 10272 (2024).
37. Yang, Y.-F. Fano effect in the point contact spectroscopy of heavy-electron materials. *Phys. Rev. B* **79**, 241107(R) (2009).
38. Maltseva, M., Dzero, M. & Coleman, P. Electron cotunneling into a Kondo lattice. *Phys. Rev. Lett.* **103**, 206402 (2009).
39. Figgins, J. & Morr, D. K. Differential conductance and quantum interference in Kondo systems. *Phys. Rev. Lett.* **104**, 187202 (2010).
40. Wolfle, P., Dubi, Y. & Balatsky, A. V. Tunneling into clean heavy fermion compounds: origin of the Fano line shape. *Phys. Rev. Lett.* **105**, 246401 (2010).
41. Shen, S. *et al.* Inducing and tuning Kondo screening in a narrow-electronic-band system. *Nat. Commun.* **13**, 2156 (2022).
42. Guillamón, I. *et al.* Chiral charge order in the superconductor 2H-TaS$_2$. *New J. Phys.* **13**, 103020 (2011).
43. Qiao, S. *et al.* Mottness Collapse in 1T−TaS$_{2-x}$Se$_x$ Transition-Metal Dichalcogenide: An Interplay between Localized and Itinerant Orbitals. *Phys. Rev. X* **7**, 041054 (2017).
44. Chen, Y. *et al.* Strong correlations and orbital texture in single-layer 1T-TaSe$_2$. *Nat. Phys.* **16**, 218-224 (2020).
45. Giannakis, I. *et al.* Orbital-selective Kondo lattice and enigmatic *f* electrons emerging from inside the antiferromagnetic phase of a heavy fermion. *Sci. Adv.* **5**, eaaw9061 (2019).
46. Cao, L. *et al.* Directly visualizing nematic superconductivity driven by the pair density wave in NbSe$_2$. *Nat. Commun.* **15**, 7234 (2024).

47. Soumyanarayanan, A. *et al.* Quantum phase transition from triangular to stripe charge order in NbSe$_2$. *Proc. Natl. Acad. Sci. USA* **110**, 1623-1627 (2013).
48. Ugeda, M. M. *et al.* Characterization of collective ground states in single-layer NbSe$_2$. *Nat. Phys.* **12**, 92-97 (2015).
49. Liu, X., Chong, Y. X., Sharma, R. & Davis, J. C. S. Discovery of a Cooper-pair density wave state in a transition-metal dichalcogenide. *Science* **372**, 1447-1452 (2021).
50. Li, H. *et al.* Rotation symmetry breaking in the normal state of a kagome superconductor KV$_3$Sb$_5$. *Nat. Phys.* **18**, 265-270 (2022).
51. Hamidian, M. H. *et al.* How Kondo-holes create intense nanoscale heavy-fermion hybridization disorder. *Proc. Natl. Acad. Sci. USA* **108**, 18233-18237 (2011).
52. Pirie, H. *et al.* Visualizing the atomic-scale origin of metallic behavior in Kondo insulators. *Science* **379**, 1214-1218 (2023).
53. Figgins, J. & Morr, D. K. Defects in heavy-fermion materials: unveiling strong correlations in real space. *Phys. Rev. Lett.* **107**, 066401 (2011).
54. Zhu, J. X., Julien, J. P., Dubi, Y. & Balatsky, A. V. Local electronic structure and Fano interference in tunneling into a Kondo hole system. *Phys. Rev. Lett.* **108**, 186401 (2012).
55. Song, Z. D. & Bernevig, B. A. Magic-Angle Twisted Bilayer Graphene as a Topological Heavy Fermion Problem. *Phys. Rev. Lett.* **129**, 047601 (2022).
56. Kumar, A., Hu, N. C., MacDonald, A. H. & Potter, A. C. Gate-tunable heavy fermion quantum criticality in a moiré Kondo lattice. *Phys. Rev. B* **106**, L041116 (2022).
57. Zhao, W. *et al.* Gate-tunable heavy fermions in a moire Kondo lattice. *Nature* **616**, 61-65 (2023).
58. Chandra, P., Coleman, P. & Flint, R. Hastatic order in the heavy-fermion compound URu$_2$Si$_2$. *Nature* **493**, 621-626 (2013).
59. Flint, R., Chandra, P. & Coleman, P. Hidden and Hastatic Orders in URu$_2$Si$_2$. *J. Phys. Soc. Jpn.* **83**, 061003 (2014).
60. Chandra, P., Coleman, P. & Flint, R. Hastatic order in URu$_2$Si$_2$: Hybridization with a twist. *Phys. Rev. B* **91**, 205103 (2015).
61. Liu, Y. *et al.* Tuning the charge density wave and superconductivity in 6$R$-TaS$_{2-x}$Se$_x$. *J. Appl. Phys.* **117**, 163912 (2015).

**Acknowledgements**
The work is supported by National Key R&D Program of China with Grants 2019YFA0307800 (J.M.), 2021YFA1600201 (X. L. and Y. P. S.), 2023YFA1607402 (X. L.), and 2022YFA1402203 (Y.-F.Y.), National Natural Science Foundation of China with Grants 12074377 (Y.J.), 11974347 (J.M.), 12321004 (Y.Y.), 12234003 (Y.Y.), W2511003 (Y.Y.), 12274412 (Y. P. S.), 12504167 (L.C.), Systematic Fundamental Research Program Leveraging Major Scientific and Technological Infrastructure, Chinese Academy of Sciences under Contract No. JZHKYPT-437 2021-08 (Y. P. S.), the China Postdoctoral Science Foundation with Grant No. 2022M723111 (L.C.), the Fellowship of China National Postdoctoral Program for Innovative Talents with grant No. BX20230358 (L.C.) and the Fundamental Research Funds for the Central Universities (L.C., J.M. and Y.J.).

**Author contributions**
Y.J., J.M. and L.C. conceived and designed the experiments. L.L., X.L., Y.S. provide the samples. L.C. performed the STM experiments. L.C., J.S., Y.-F. Y., Y. Y., J.M. and Y.J. analyzed the raw data. L.C. and J. S. plotted the figures. L.C., J.S., Y.-F. Y., Y. Y., J.M. and Y.J. wrote the manuscript with inputs from all authors. Y.J., J.M. and Y. Y., supervised the project.

**Competing interests**
The authors declare no competing interests.

**Additional information**
**Correspondence and requests for materials** should be addressed to Yugui Yao, Jinhai Mao and Yuhang Jiang

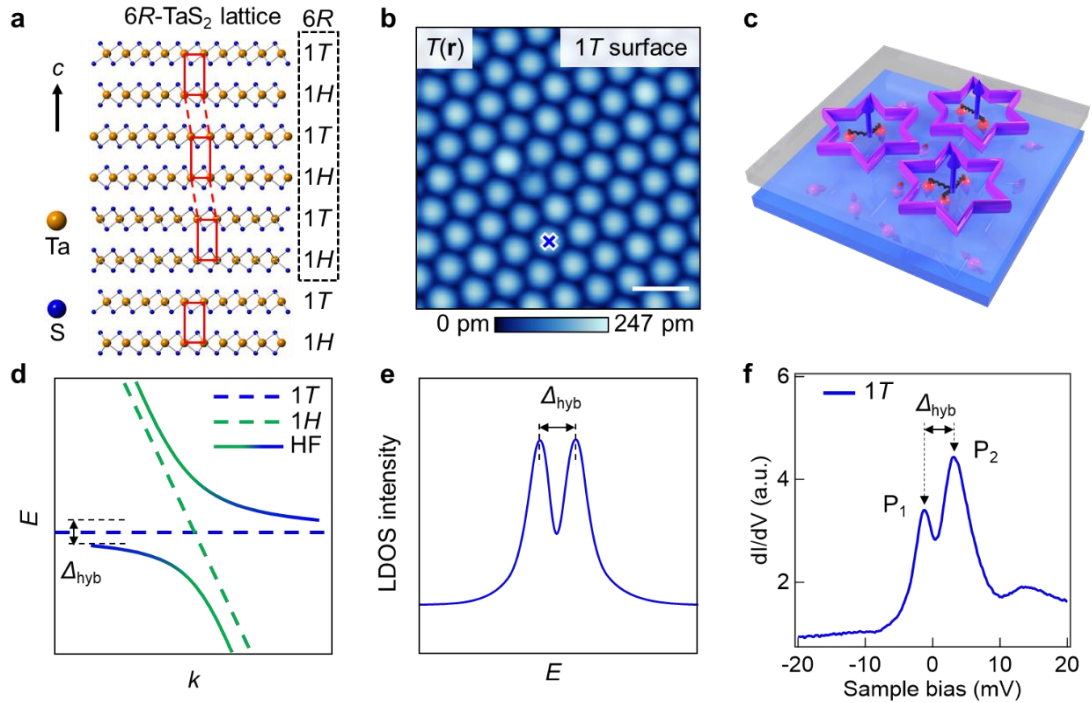

**Fig. 1 | Kondo lattice and hybridization gap in 6$R$-TaS$_2$. a**, The lattice structure of 6$R$-TaS$_2$, where 1$T$-TaS$_2$ layers and 1$H$-TaS$_2$ layers stack alternately along the $c$ axis. Red rectangles mark every 1$T$/1$H$ bilayers and red dashed lines indicate the in-plane glide symmetry. Black dashed rectangle indicates one full cycle (6$R$). **b**, The STM topography image $T(\mathbf{r})$ of the 1$T$ surface. The Star-of-David (SD) superlattice with period $a_{SD}$ = 1.18 nm are resolved. Scale bar: 2 nm. **c**, Schematic of the Kondo coupling between the local magnetic moments from 1$T$ layer and itinerant electrons from 1$H$ layer. **d**, Schematic of band structures for 1$T$ and 1$H$ layers and the heavy fermion (HF) band with hybridization gap ($\Delta_{hyb}$). **e**, Schematic of the LDOS when tunneling into the local magnetic moment is dominant. **f**, The d$I$/d$V$ spectrum measured at the cross on 1$T$ surface (**b**), showing two hybridization peaks P$_1$ and P$_2$ and the hybridization gap $\Delta_{hyb}$, which is the hallmark of Kondo lattice. Setpoints for (**b**): $V_s$ = –0.1 V $I_t$ = 60 pA, for (**f**): $V_s$ = –30 mV and $I_t$ = 200 pA.

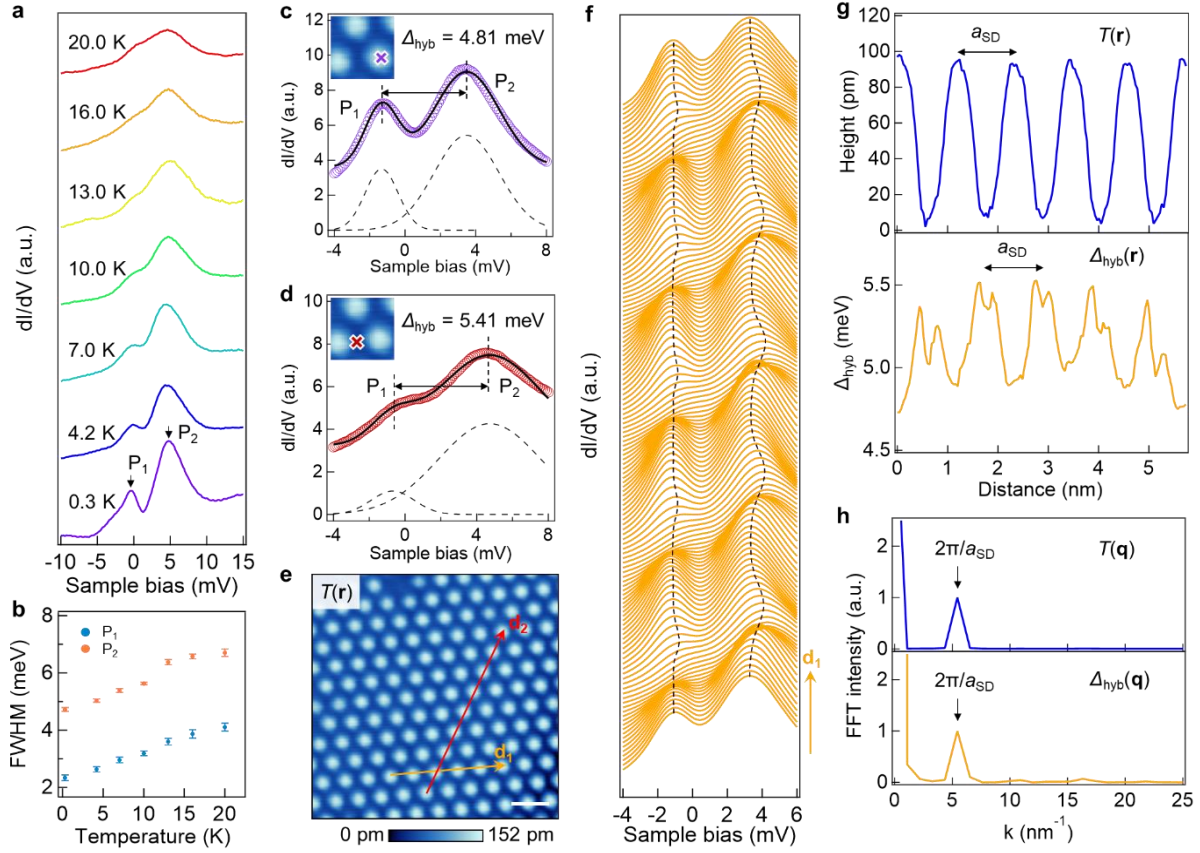

**Fig. 2 | Temperature and Spatial evolution of hybridization peaks. a**, Temperature dependent d$I$/d$V$ spectra for hybridization peaks P$_1$ and P$_2$. **b**, The extracted full width at half maximum (FWHM) for P$_1$ and P$_2$ as the function of temperature. **c**, The d$I$/d$V$ spectrum (purple circles) taken at SD center, as marked in the inset. The black solid curve is the multiple-peak fitting result by two Gaussian peaks (dashed curves). $\Delta_{hyb}$ = 4.81 meV at SD center. **d**, Same of (**c**) but taken at SD edge. $\Delta_{hyb}$ = 5.41 meV at SD edge. **e**, The topography $T(\mathbf{r})$ of region #1. Arrows for $\mathbf{d}_1$ and $\mathbf{d}_2$ mark where the one-dimensional d$I$/d$V(\mathbf{r}, V)$ spectra maps are taken. Scale bar: 2 nm. **f**, Water-fall plot of the one-dimensional d$I$/d$V(\mathbf{r}, V)$ spectra map measured along $\mathbf{d}_1$, as indicated by the orange arrow in (**e**). Spatial dependence of hybridization peaks' energy $E_{P1}(\mathbf{r})$ and $E_{P2}(\mathbf{r})$ are tracked by the dashed lines. **g**, Topography $T(\mathbf{r})$ (upper panel) and spatial dependent $\Delta_{hyb}(\mathbf{r})$ (lower panel) extracted from dataset in (**f**). **h**, The FFT results of (**g**), i.e., $T(\mathbf{q})$ and $\Delta_{hyb}(\mathbf{q})$. Setpoints for (**a**), (**c**), (**d**), (**e**), (**f**): $V_s$ = –30 mV, $I_t$ = 200 pA.

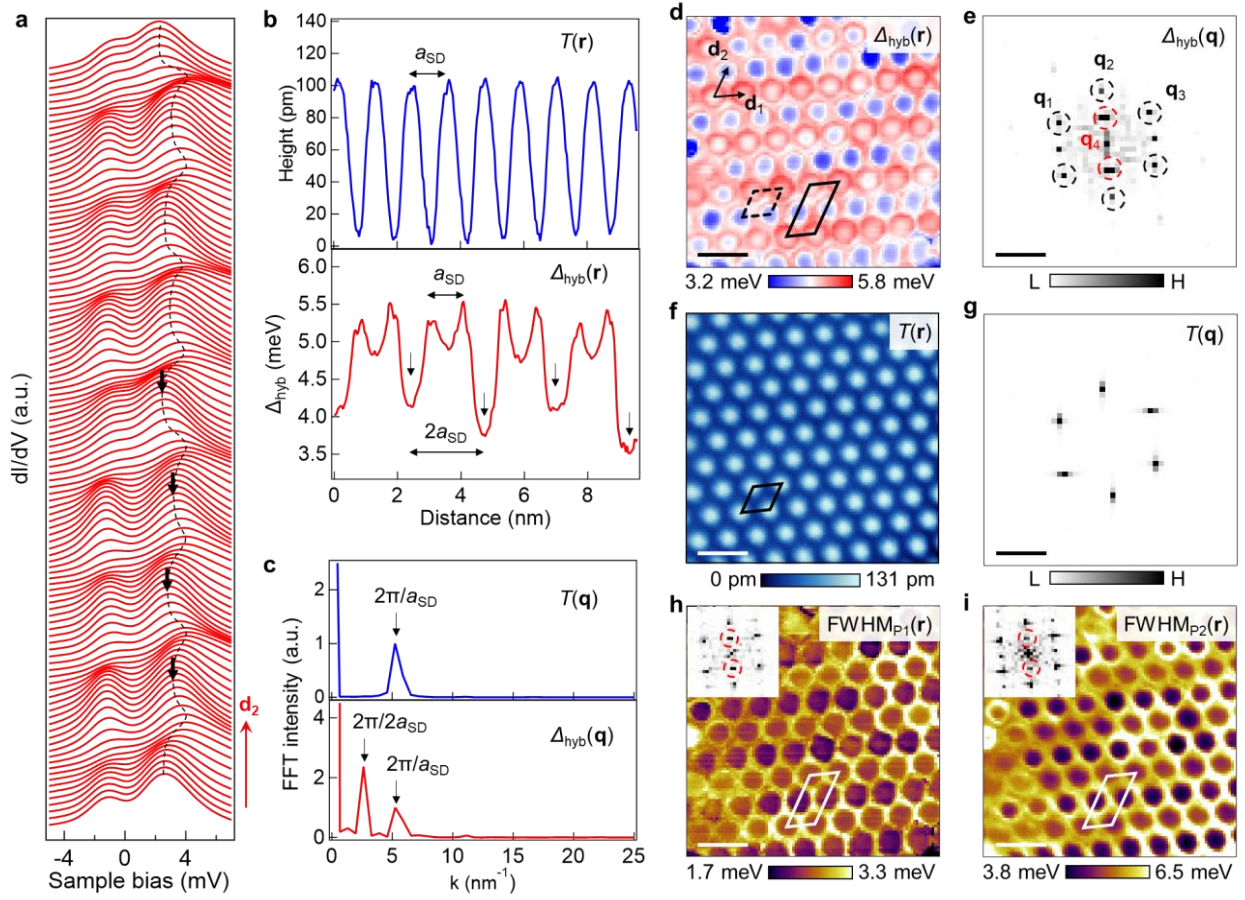

**Fig. 3 | Evidence for hybridization wave. a**, Water-fall plot of the one-dimensional d$I$/d$V$(**r**, $V$) spectra map measured along **d**$_2$, as indicated by the red arrow in Fig. 2e. The different energy locations for P$_2$ at two adjacent SD centers are highlighted by black arrows. **b**, Topography $T(\mathbf{r})$ (upper panel) and spatial dependent $\Delta_{\text{hyb}}(\mathbf{r})$ (lower panel) extracted from dataset in (**a**). Vertical arrows highlight the unit-cell doubling behavior of $\Delta_{\text{hyb}}$ with period of 2$a_{\text{SD}}$. **c**, The FFT results of (**b**), i.e., $T(\mathbf{q})$ and $\Delta_{\text{hyb}}(\mathbf{q})$. The wave vector at $2\pi/2a_{\text{SD}}$ which corresponds the unit-cell doubling emerges. **d**, two-dimensional hybridization gap map $\Delta_{\text{hyb}}(\mathbf{r})$ in region #1. The dashed diamond is the SD unit-cell and the solid parallelogram highlights the unit-cell doubling. **e**, FFT image of (**d**), i.e., $\Delta_{\text{hyb}}(\mathbf{q})$. Black dashed circles indicate the wave vector of $2\pi/a_{\text{SD}}$ at three high symmetric directions. Red dashed circles highlight the uniaxial wave vector at $2\pi/2a_{\text{SD}}$. **f**, The corresponding topography $T(\mathbf{r})$ of (**d**). The solid diamond marks the SD unit-cell. **g**, FFT image of (**f**). **h-i**, The real-space FWHM map for P$_1$ and P$_2$, respectively. The solid parallelograms highlight the unit-cell doubling. Insets: the FFT images of the FWHM maps. Red dashed circles highlight the uniaxial wave vector at $2\pi/2a_{\text{SD}}$. Scale bars in (**d**), (**f**), (**h**) and (**i**): 2 nm, in (**e**) and (**g**): 6.10 nm$^{-1}$. Setpoints for (**a**), (**d**), (**f**), (**h**) and (**i**): $V_s = -30$ mV, $I_t = 200$ pA.

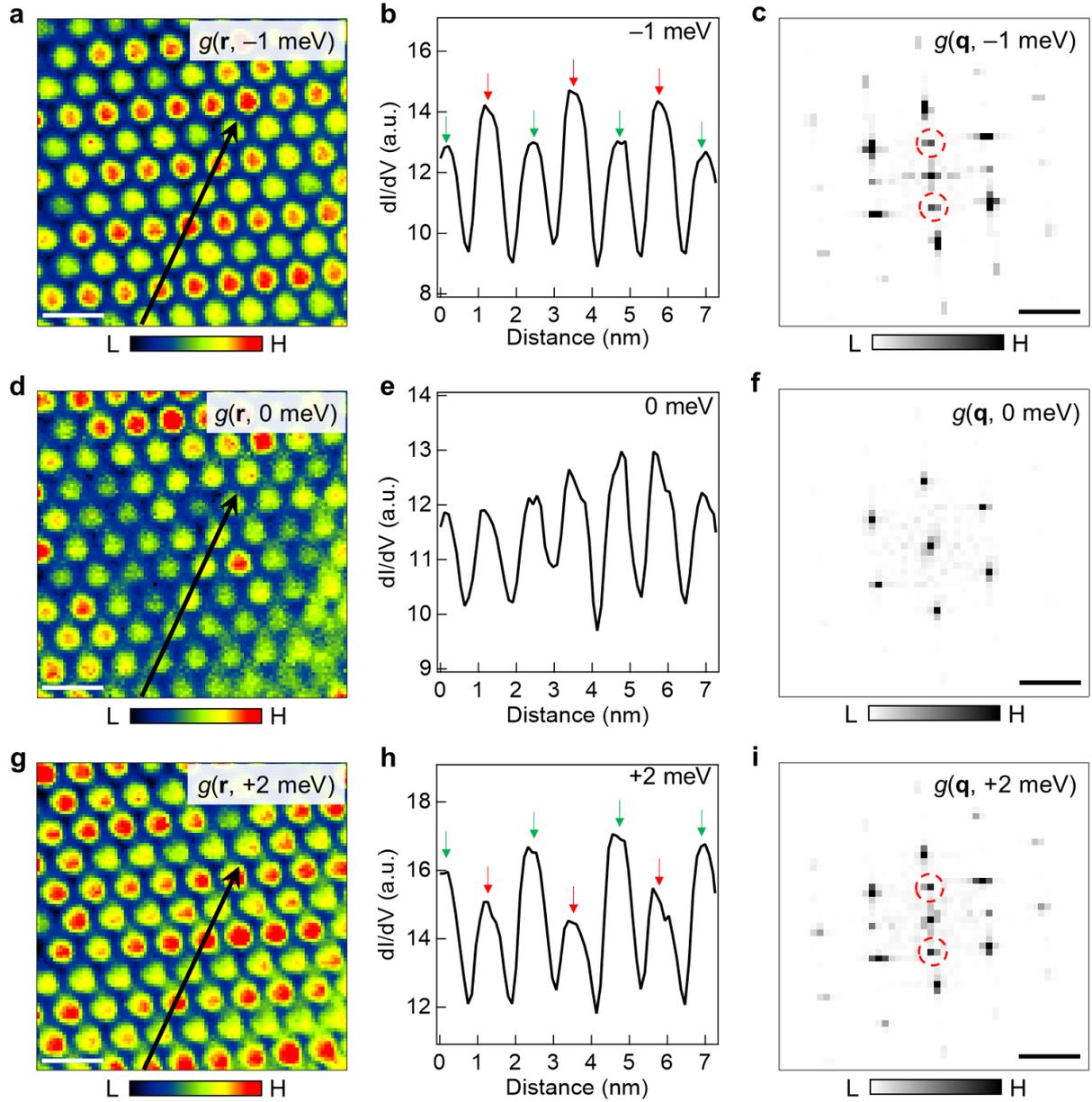

**Fig. 4 | Energy-dependent nematic order. a**, The LDOS maps $g(\mathbf{r}, E)$ measured at energy $E = -1$ meV. Nematic order is clearly resolved. **b**, The line-profile extracted along black arrow in (**a**), crossing several SD centers. The spatial locations of relatively high and low LDOS intensity at each SD center are marked by red and green arrows, respectively. **c**, The FFT image of (**a**). Wave vector of the nematic order is highlighted by the red dashed circles. **d-f**, Same of (**a**)-(**c**) but for energy $E = 0$ meV. The nematic order disappears at this energy in both real-space and reciprocal-space. **g-i**, Same of (**a**)-(**c**) but for energy $E = +2$ meV. The nematic order reappears at this energy. The spatial locations of relatively high and low LDOS intensity at each SD center are marked by green and red arrows respectively. Scale bars in (**a**), (**d**), (**g**): 2 nm, in (**c**), (**f**), (**i**): 6.10 nm$^{-1}$. Setpoints for (**a**)-(**c**): $V_s = -30$ mV, $I_t = 200$ pA.